\documentclass[a4paper,conference]{IEEEtran}
\usepackage[ruled,vlined,linesnumbered]{algorithm2e}
\usepackage{amsmath}
\usepackage{mathtools}
\usepackage{amsthm}
\DeclarePairedDelimiter{\ceil}{\lceil}{\rceil}

\theoremstyle{remark}

\theoremstyle{definition}

\newcommand{\E}{\mathrm{E}}

\title{Random Access for Machine-Type Communication based on Bloom Filtering}
\author{\IEEEauthorblockN{Nuno K. Pratas, \v Cedomir Stefanovi\'c, Germ\'an Corrales Madue\~no, Petar Popovski} \\
\IEEEauthorblockA{Department of Electronic Systems, Aalborg University, Denmark\\
Email: \{nup,cs,gco,petarp\}@es.aau.dk}%
}

\begin{document}
\maketitle
\begin{abstract}
We present a random access method inspired on Bloom filters that is suited for Machine-Type Communications (MTC).
Each accessing device sends a \emph{signature} during the contention process.
A signature is constructed using the Bloom filtering method and contains information on the device identity and the connection establishment cause.
We instantiate the proposed method over the current LTE-A access protocol.
However, the method is applicable to a more general class of random access protocols that use preambles or other reservation sequences, as expected to be the case in 5G systems.
We show that our method utilizes the system resources more efficiently and achieves significantly lower connection establishment latency in case of synchronous arrivals, compared to the variant of the LTE-A access protocol that is optimized for MTC traffic.
A dividend of the proposed method is that it allows the base station (BS) to acquire the device identity and the connection establishment cause already in the initial phase of the connection establishment, thereby enabling their differentiated treatment by the BS.
\end{abstract}

\IEEEpeerreviewmaketitle

\section{Introduction} 
\label{sec:introduction}

\begin{figure}[t]
	\centering
		\includegraphics[width=0.9\linewidth]{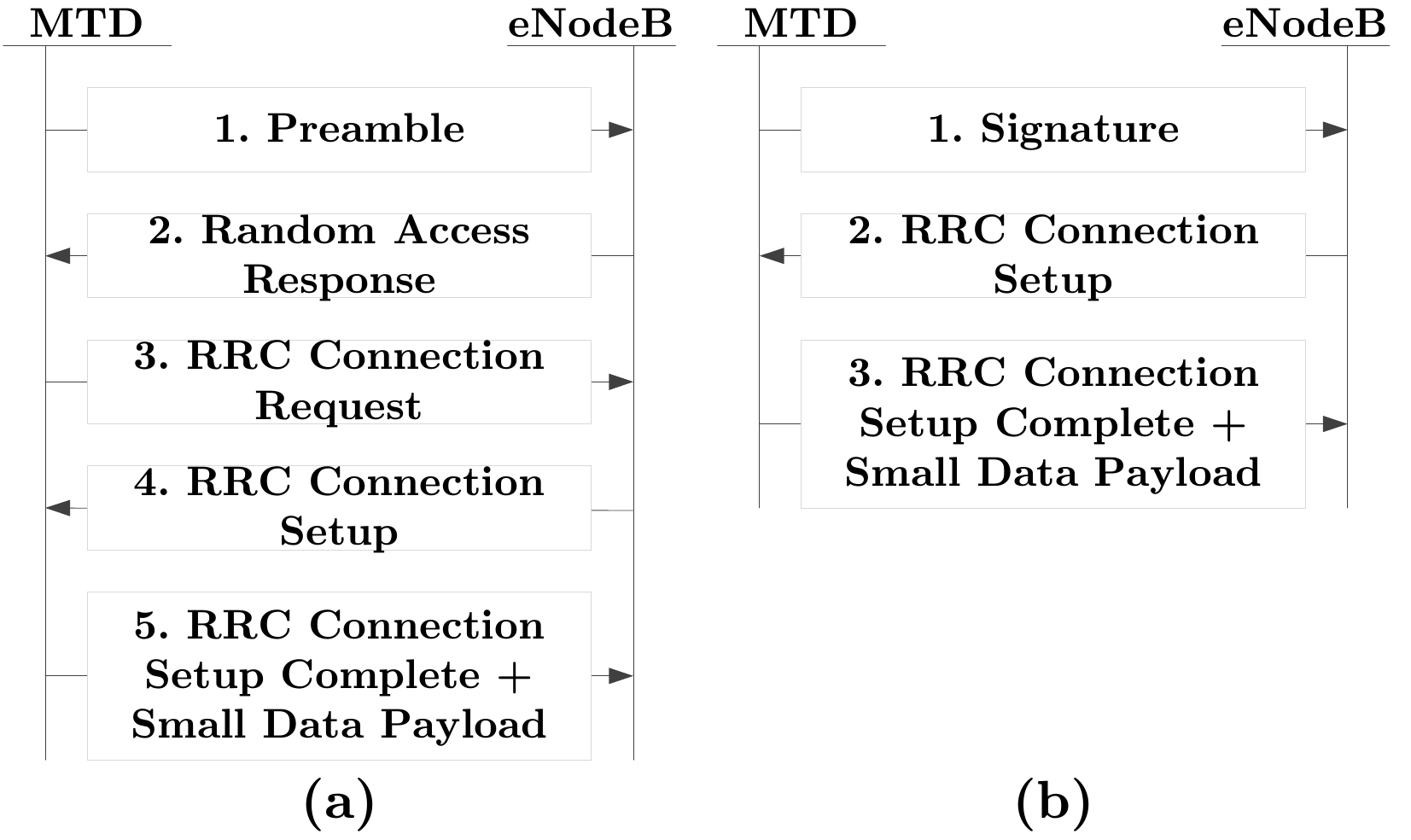}
	\caption{(a) LTE-A connection establishment protocol optimized for MTC~\cite{3GPPTR36.888} and (b) signature-based modification of LTE-A connection establishment.}
	\vspace{-0.5cm}
	\label{fig:ARPComparison}
\end{figure}

Machine-type communications (MTC) are typically characterized by a massive number of machine-type devices that connect to the network to transmit small data payloads.
Those features present a significant challenge to cellular networks, whose radio access part is traditionally designed to deal with a rather low number of connections with high data requirements.
Specifically, current cellular networks, such as LTE-A, are connection-oriented~\cite{TribudiWiriaatmadja2014}, requiring a connection establishment between the device and the Base Station (BS) before the device can transmit its data packet.
As an example, the connection establishment in LTE-A involves a high amount of signaling overhead, which is particularly emphasized when the data payload is small, e.g., less than 1000 bytes~\cite{3GPPTR37.869}.
Therefore, in 3GPP it was proposed an approach to optimize the connection establishment by reducing the signaling overhead~\cite{3GPPTR36.888}. 
The resulting simplified connection establishment protocol starts with the contention-based Access Reservation Protocol (ARP)~\cite{3GPPTS36.321}, depicted in the first four steps in Fig.~\ref{fig:ARPComparison}(a), followed by a fifth message where the signaling and a small data payload are concatenated.
The signaling exchanges related to the security mechanisms are omitted in the optimized version of the LTE-A connection establishment, by reusing an a-priori established security context~\cite{3GPPTR37.869}.

The throughput and blocking probability of the ARP are rather sensitive to the number of contending devices.
Specifically, the devices contend for access by sending their preambles in a designated and periodically occurring uplink sub-frame, here termed as random access opportunity (RAO).
When the number of contending devices is high~\cite{7397849}, multiple devices activate the same preamble in a RAO, which leads to collisions of their RRC Connection Requests, see Fig.~\ref{fig:ARPComparison}(a).
Consequently, most devices are unable to establish a connection in the first attempt and perform subsequent attempts that, due to the high load, are also likely to result in collisions.
A solution put forward to cope with congestion, was the extended access class barring (EAB) \cite{36331}, where certain classes of devices are temporally blocked from participating in the ARP, but at the cost of an increased access latency to those same devices.
Another drawback of the ARP is that the network learns the devices' identities and connection establishment causes only after the RRC Connection Request is successfully received, as the contention is performed via randomly chosen preambles that do not carry information.
A solution that allows the network to learn the identities and connection establishment causes of the contending devices already at the beginning of the ARP, could enable their differentiated treatment in later phases of the connection establishment and even skip some of the steps in the LTE-A random access protocol, as indicated in Fig.~\ref{fig:ARPComparison}.

In this paper we propose a new access method based on signatures and Bloom filtering~\cite{Bloom1970}.
The method is demonstrated in the context of the LTE-A ARP, however, we note that it can be employed in the next generation ARPs~\cite{FANTASTICIR412016} following similar principles.
In the proposed method, instead of contending with a single preamble in a RAO, the devices contend by transmitting a predefined sequence of preambles in a frame composed of several RAOs, 
The transmitted sequence of preambles is denoted as the \emph{device signature}.
The presented ideas are a conceptual extension of the work \cite{ETT:ETT2656}, where the devices contend for access by selecting a random signature, generated by combining random preambles over consecutive RAOs.
In contrast, in the method described here, each device contends with a unique signature generated using the International Mobile Subscriber Identity (IMSI) of the device and its connection establishment cause, in further text referred to as the device's identification.\footnote{We note that the proposed method can be straightforwardly applied to cases where some other information is used for signature generation.}
Specifically, we apply the Bloom-filter~\cite{Bloom1970} principles for signature generation, where the device's identification is hashed over multiple independent hash functions and the resulting output used to select which preamble in which RAO to activate.
We introduce an analytical framework through which we tune the signature properties, i.e., its length and the number of activated preambles, based on the number of expected arrivals and the target efficiency of the use of system resources, denoted as the goodput.
We also investigate the expected latency and signature detection probability of the proposed method.
Finally, we show that, when the arrivals are synchronous, the proposed method outperforms the LTE-A connection establishment procedure in terms of goodput, while achieving similar or lower average latency.

The rest of the paper is organized as follows. 
Section~\ref{sec:LTE_ARP} summarizes the standard ARP in LTE-A.
Section~\ref{sec:proposed_contention_modifications} describes the proposed access method and Section~\ref{sub:analytical_performance_model} presents the corresponding analysis.
Section~\ref{sec:system_performance_evaluation} evaluates the performance of the proposed method, comparing it with the reference LTE-A procedure for MTC traffic.
Section~\ref{sec:conclusions} concludes the paper.


\section{LTE-A Access Reservation Procedure}
\label{sec:LTE_ARP}

A successful LTE-A access reservation entails the exchange of four messages\footnote{For the sake of brevity, we omit the details that are nonessential for the proposed method, such as the power ramping procedure etc.}, as depicted in Fig.~\ref{fig:ARPComparison}(a).
Initially, a device randomly chooses a preamble to be transmitted in a RAO from a set of available preambles generated using Zadoff-Chu sequences~\cite{1054840}.
The preambles are orthogonal and can be simultaneously detected by the BS.
We also note that the BS is able to detect a preamble even when it is transmitted by multiple devices~\cite{TribudiWiriaatmadja2014,ETT:ETT2656}, i.e., a collision in the ``preamble space'' is still interpreted as an activated preamble.
This represents a logical OR operation, since the preamble is detected as activated if there is \emph{at least} one device that transmits the preamble.
This observation motivates the use of Bloom filter, a data structure based on OR operation for testing set membership.

The devices whose preambles are detected are notified via a Random Access Response (RAR) in the downlink and assigned a temporary network identifier.
The reception of the RAR triggers the transmission of the RRC Connection Request in the allocated uplink sub-frame.
At this point, the BS is able to detect the collision of the multiple connection requests, sent by the devices that originally sent the same preamble.
The successfully received connection requests are acknowledged, marking the start of the data transmission phase.
On the other hand, the devices whose connection requests collided, do not receive the feedback and either contend again by sending a new preamble or end up in outage when the number of connection attempts reaches the predefined limit.
In the RRC Connection Request, the device informs the network of its temporary identifier, IMSI, and the connection establishment cause.
From these, the network can confirm if the device is authorized for access, track the device's subscribed services and reestablish the preexisting security context~\cite{3GPPTR37.869}.

As already mentioned, the channel over which the devices contend can be modeled as an OR multiple access channel (OR-MAC).
By $A=\{a_i, i = 0,1,..., M \}$, denote the set of available preambles, where the absence of preamble activation is denoted by the idle preamble $a_0$.
Assume that there are $T$ devices in total.
We model the contention by assuming that the device $h$, $h=1,\dots, T$, transmits a binary word
\begin{align}\label{eq:x}
	\mathbf{x}^{(h)} = [ x^{(h)}_0, x^{(h)}_1, \cdots, x^{(h)}_M ],
\end{align}
where bit $x^{(h)}=1$ indicates if the device $h$ transmitted preamble $a_i$. Note that only a single entry $x^{(h)}_i$, $0 \leq i \leq M$, can be set to 1 since a device can only transmit a single preamble in a single RAO.
The BS observes
\begin{align}\label{eq:y}
	\mathbf{y} = \bigoplus_{h = 1}^{T} \hat{\mathbf{x}}^{(h)},
\end{align}
where $\bigoplus$ denotes a bit-wise OR operator and $\hat{\mathbf{x}}^{(h)}$ is the detected binary word of device $h$.
In particular, the BS detects a transmitted preamble with probability $p_d \leq 1$ and with probability $p_f \geq 0$ falsely detects a non-transmitted preamble, which may cause that $\mathbf{x}^{(h)} \neq \hat{\mathbf{x}}^{(h)}$.
In practice, the preamble detection at the BS should ensure that $p_d > 0.99$ and $p_f <10^{-3}$~\cite{3GPPTS36.141}\footnote{The $p_d$ requirement in~\cite{3GPPTS36.141} corresponds to the single activation of a preamble.
When a preamble is activated by multiple devices it is expected that the effective $p_d$ will be higher~\cite{TribudiWiriaatmadja2014}.}.
Finally, every non-zero entry in $\mathbf{y}$ implies a detection of the corresponding preamble.
Obviously, in the best-case scenario, the BS can detect up to $M$ different devices in a RAO. 

\section{The Proposed Method} 
\label{sec:proposed_contention_modifications}
\begin{figure}[t]
	\centering
	\includegraphics[width=\linewidth]{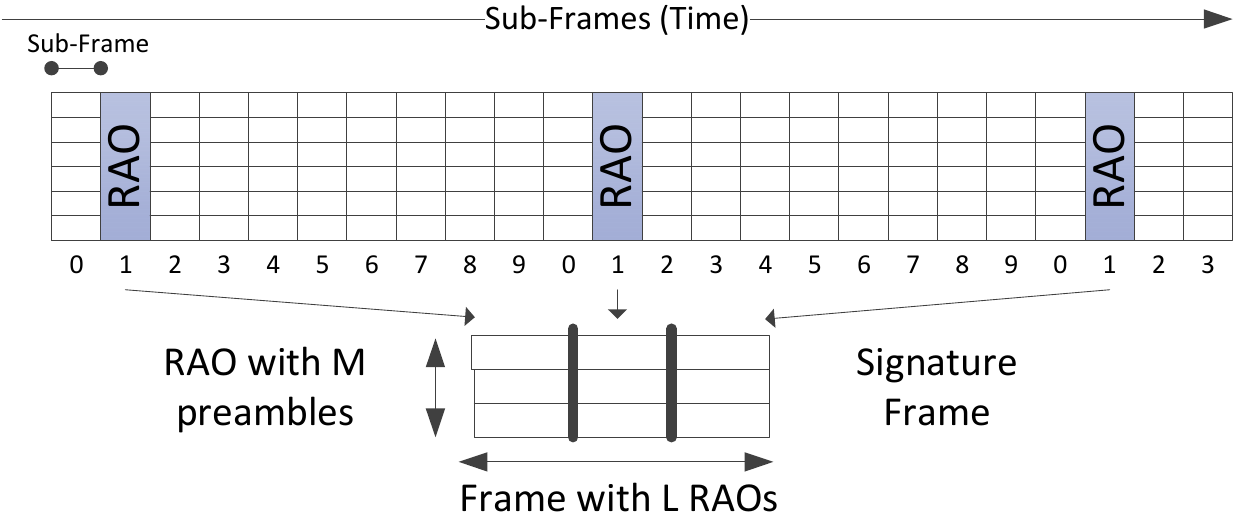}
	\vspace{-0.5cm}
	\caption{Illustration of the mapping of the LTE-A preambles into a signature frame composed by multiple RAOs.}
	\vspace{-0.5cm}
	\label{fig:LTEORMAC}
\end{figure}
The essence of the proposed method lies in the idea of devices contending with combinations of $K$ preambles transmitted over $L$ RAOs, denoted as signatures.
Each preamble of a signature is sent in a separate RAO, while $L$ RAOs define a signature frame, see Fig.~\ref{fig:LTEORMAC}.
Extending the model introduced in Section~\ref{sec:LTE_ARP}, the device $h$ contends by transmitting its signature
\begin{align}
	\mathbf{s}^{(h)} = [ \mathbf{x}^{(h)}_{1}, \mathbf{x}^{(h)} _{2}, \cdots, \mathbf{x}^{(h)} _{L}], 
\end{align}
where the binary words $\mathbf{x}^{(h)}_i$, $i = 1, \dots, L$, follow the structure introduced in \eqref{eq:x}.
Obviously, the number of available signatures is $\binom{L}{K} M^K$, potentially allowing for the detection of exponentially more contenders compared to the case in which the preambles sent in each of the $L$ RAOs are treated independently and where the maximal number of detected contenders is $L \cdot M$.

Similarly to \eqref{eq:y}, the BS observes
\begin{equation}\label{eq:y_new}
	\mathbf{y} = \bigoplus_{h = 1}^{N} \hat{\mathbf{s}}^{(h)},
\end{equation}
where $\hat{\mathbf{s}}^{(h)}$ is the detected version of $\mathbf{s}^{(h)}$.
The BS decodes all signatures $\mathbf{s}$ for which the following holds
\begin{align}\label{eq:det}
	\mathbf{s} = \mathbf{s} \bigotimes \mathbf{y},
\end{align}
where $\bigotimes$ is the bit-wise AND. 

\begin{figure}[t]
	\centering
	\includegraphics[width=\linewidth]{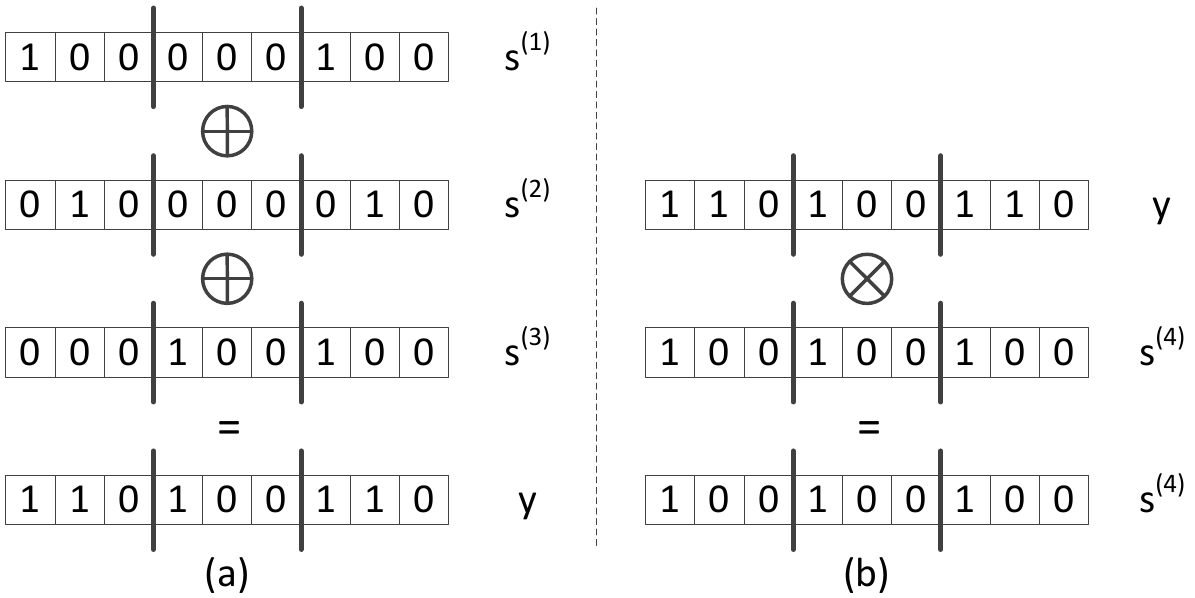}
	\caption{Example of: (a) synchronous transmission of $3$ signatures when $L = 3$ and $M = 3$ and (b) erroneous decoding of a signature which was not present in the original transmission ($p_d = 1$ and $p_f = 0$).}
	\vspace{-0.5cm}
	\label{fig:ORMACSignatureTransmissionDetectionExample}
\end{figure}

At this point, we turn to a phenomenon intrinsically related to the proposed contention method~\cite{ETT:ETT2656}.
Namely, even in the case of perfect preamble detection ($p_d = 1$) and no false detections ($p_f = 0$), the BS may also decode signatures that have \emph{not} been transmitted but for which \eqref{eq:det} also holds.
In other words, the BS may decode \emph{false positives}.
An example of this is shown in Fig.~\ref{fig:ORMACSignatureTransmissionDetectionExample}.
The performance of the random signature construction in terms of probability of decoding false positives was first analyzed in \cite{ETT:ETT2656}, where they are referred to as phantom sequences.
On the other hand, there is an extensive work on the construction of OR-MAC signatures~\cite{Gyori20081407} based on the following criterion: if up to $N$-out-of-$T$ signatures are active, then there are no false positives.
However, these constructions are not directly applicable to the LTE-A access, as they would (1) require that a device sends multiple preambles in the same RAO, and (2) imply rather long signature lengths, i.e., $\frac{N^2 \log_2 T}{2 M \log_2 N} \leq L \leq \frac{N^2 \log_2 T}{M\ln 2}$, which implies an increased access latency.
Inspired by Bloom filters~\cite{Bloom1970}, we propose a novel signature construction that uses much lower signature lengths, at the expense of introducing false positives in a controlled manner.

\subsection*{Signature Construction based on Bloom Filtering} 
\label{sub:bloom_filter_inspired_signatures}

In the proposed method, the device signature is constructed in such a way that it provides a representation of the device's identification, which is assumed to be a-priori known to the network. 
To illustrate how a signature is constructed, we first consider the case where a single preamble is available at each of the $L$ RAOs dedicated to the signature transmission, i.e., $M=1$.
Taking the view of the device $h$, we start with the binary array $\mathbf{s}^{(h)}$ of length $L$, indexed from $1$ to $L$, where all the bits are initially set to $0$.
We then activate $K$ index positions in this array, i.e., we set them to $1$; note that $K$ is a predefined constant valid for all devices.
This is done by using $K$ independent hash functions, $f_j ( \mathbf{u}^h )$, $j = 1, \dots, K$, whose output is an integer value between 1 and $L$, corresponding to an index position in the array, and where $\mathbf{u}^{(h)}$ is representation of the device identity.
The resulting binary array becomes the device signature.
This construction follows the same steps as the object insertion operation in a Bloom filter~\cite{Bloom1970}.

When $M>1$, the signature construction occurs in two stages.
The first stage corresponds to the selection of the $K$ active RAOs using hash functions $f_j ( \mathbf{u}^h )$, $j = 1, \dots, K$, as described previously.
In the second stage, for each of the activated RAOs, a contending device selects and transmits randomly one of $M$ preambles.
This is performed by hashing the device identity using another set of independent hash functions $g_j ( \mathbf{u}^h )$, $j = 1, \dots, K$, i.e., a separate hash function for each RAO, whose output is an integer between $1$ and $M$ that corresponds to one of the available preambles.

\subsection*{Signature-Based ARP} 
\label{sub:signature_ARP}

The signature-based access reservation protocol is depicted in Fig.~\ref{fig:ARPComparison}(b), which starts by the devices transmitting their signatures.
Upon the successful decoding of a signature, the BS transmits the \emph{RRC Connection Setup} message.
In contrast with the LTE-A ARP depicted in Fig.~\ref{fig:ARPComparison}(a), the messages 2 and 3 are not required in the signature based access, since the BS is able to determine from the signature the IMSI of the device and the connection establishment cause.
The protocol concludes with the transmission of the small data payload together with the completion of the RRC connection message.

\subsection*{Practical Considerations} 
\label{sub:practical_considerations}

The described signature generation raises two important issues: (i) out of $K$ hash functions $f_j ( \mathbf{u}^h )$,  $j = 1, \dots, K$, there is a probability of $1 - K!\binom{L}{K}/L^K$ that at least two of these functions generating the same output, leading to less than $K$ distinct RAOs active in a signature; (ii) there is a non-zero probability that two or more devices share the same signature, given by
\begin{equation}
	\sum_{i=2}^{T} \binom{T}{i} p^i(1-p)^{T-i} \mbox{ with } p = \left[ \binom{L}{K} (M)^K \right]^{-1}
\end{equation}
and $T$ as the total number of devices.
The above probabilities can be minimized by increasing the signature length $L$, which is the reason why these issues are commonly ignored within the Bloom filter related literature, where $L$ is of the order of $10^4$.
Although we do not use such large ranges for $L$, we note that for values of $L>10$ and $5 < K < L$ that are used in the performance evaluation in Section~\ref{sec:system_performance_evaluation}, the second probability can be neglected, as in this case $T \ll \binom{L}{K} (M)^K$.
\begin{algorithm}[t]\label{alg:bloomfilterinsertion}
	\textbf{Input}: {$\mathbf{u}^{(h)}$, $L$, $M$, $K$}; \\
	\textbf{Initialize}: $\mathbf{s}^{(h)} \gets \mathbf{0} $, $ \mathbf{L} \gets 1...L$, $ \mathbf{M} \gets 1...M$ \;
	\For{$ j : 1 \cdots K$}{
		$i \gets \mathbf{L} (\text{mod}(\mathbf{u}^{(h)},L+1-j))$\; $\mathbf{L} = \mathbf{L} \setminus \{i\}$\;
		$m \gets \mathbf{M} (\text{mod}(\mathbf{u}^{(h)},M+1-j))$\; $\mathbf{M} = \mathbf{M} \setminus \{m\}$\;
		$x_{i,m}^{(h)} = 1$\;
	}
	Output {$\mathbf{s}^{(h)}$}; \\
	\caption{Signature generation for $h^{th}$ device, where $\mathbf{u}^{(h)}$ is the device's identification and $x_{i,m}^{(h)}$ indicates activation of $m^{th}$ preamble in $i^{th}$ RAO of the signature $\mathbf{s}^{(h)}$.}
\end{algorithm}

The first issue can be addressed by a signature construction that enforces $K$ distinct active RAOs per signature.
We provide in Alg.~\ref{alg:bloomfilterinsertion} a description of a practical signature construction that uses the modulus operation as basis for the hashing.
This construction ensures that $K$ distinct RAOs are active per signature, by removing the RAOs selected in previous iterations from the set of available RAOs.
Further, the preambles activated in previously selected RAOs are removed from the set preambles available for the next iteration.
This operation limits the generation of signatures to $K\leq \min(M,L)$ active RAOs; however, this is within the operating range of interest where $K<M$ and allows us to apply probabilistic tools, as presented in the analysis in Section~\ref{sub:analytical_performance_model}, to design the signatures length $L$ and number of active RAOs $K$.
As it will be shown in Section~\ref{sec:system_performance_evaluation}, the proposed signature generation algorithm matches well the derived analytical model.

Finally, we note that an essential prerequisite for the proposed signature access scheme is that the signature generation algorithm and all the hash functions are known to all devices, including the BS. 
This can be accomplished via the existing periodic broadcasts that include the network configuration; an alternative would be to include this information already in the device's subscriber identity module.


\section{Analysis} 
\label{sub:analytical_performance_model}

We analyze a single instance of the contention process, assuming a synchronous batch arrival of $N_\text{a}$ devices.
We assume that the probability of an arrival of a device is $p_a = \E [ N_\text{a} ] /T$, and denote the expected number of arrivals as $N = \E [ N_\text{a} ]$.
The parameters of the proposed scheme are the signature frame size, denoted by $L$, the number of active RAOs in the signature, denoted by $K$, and the number of preambles per RAO that are available for signature construction, denoted by $M$.
The first two parameters are subject to design, and we analyze their dimensioning when on average $N$-out-of-$T$ signatures are active, such that the false positive rate is below a  threshold.
In contrast, $M$ is assumed to be fixed, which corresponds to the typical scenario in LTE-A systems.

We start by establishing the relationship between the correctly detected signatures and all detected signatures, which also includes the false positives, after all the contenders have completed $3^{rd}$ step of the proposed method, see Fig.~\ref{fig:ARPComparison}(b).
We denote this metric as the goodput $G$.
In essence, the goodput reflects the efficiency of the subsequent small data transmission, as the BS will also attempt to serve the falsely detected signatures.
The expected goodput is
\begin{equation}
	\label{eq:G_def}
	\E \left[ G \right] = \E \left[ \frac{N_\text{a} }{N_\text{a} + P} \right] \approx \frac{\E [N_\text{a}] }{ \E [N_\text{a}] + \E[P]} = \frac{N}{N + \E[P]}.
\end{equation}
where $P$ is the number of false positives.
From \eqref{eq:G_def} it follows
\begin{align}
	\label{eq:G_bounds}
	\frac{N}{T} \leq \E [ G ] \leq 1,
\end{align}
as there can be no more than $T$ detected signatures. 
The mean number of false positives $E[P]$ can be approximated as
\begin{equation*}
	E[P] \approx p_\text{fa}  (T - N ),
\end{equation*}
where $T- N$ corresponds to the mean number of inactive signatures, while $p_\text{fa}$ denotes the false positive probability, i.e., the probability of an inactive signature being perceived as active.
Eq. \eqref{eq:G_def} now becomes
\begin{align}
	\label{eq:G_approx}
	\E \left[ G \right] \approx \frac{N}{N + p_\text{fa} (T - N )}.
\end{align}

Using \eqref{eq:G_approx}, we proceed by setting the target goodput $\hat{G}$ and establishing the relation between $\hat{G}$ and the corresponding target $\hat{p}_\text{fa}$
\begin{equation} \label{eq:phantomTarget}
	\hat{p}_\text{fa} = \frac{N ( 1 - \hat{G}) }{ ( T - N ) \hat{G}}.
\end{equation}
To compute $p_\text{fa}$, we rely on approximations that hold when the number of simultaneously active signatures $N$ is high enough.
Specifically, $p_\text{fa}$ is the probability that all $K$ preambles associated with an inactive signature, are detected as activated by the BS.
Each of these $K$ preambles can be (i) actually activated by an active signature and detected as such by the BS, or (ii) not activated by any of the active signatures, but falsely detected as activated by the BS.
Now, the probability that a particular preamble in a particular RAO is not activated by any of the signatures, denoted by $p_\text{idle}$, is
\begin{equation}
	p_\text{idle} = \left( 1 - \frac{K}{L \cdot M}\right)^{N},
\end{equation}
where $L \cdot M$ is the total number of preambles in $L$ RAOs, $K$ is the number of preamble activations per user, $N$ is the number of active signatures, and it is assumed that the selection of any preamble in any RAO is equally likely.
The detection of a preamble is non-ideal and therefore we have to distinguish between two events: (i) detection of a preamble transmitted by at least one device with probability $p_d$; (ii) false detection of a non-transmitted preamble with probability $p_f$.
We approximate $p_\text{fa}$ as
\begin{align}\label{eq:PhantomSignature}
	p_\text{fa} &\overset{(a)}{\approx} \left[ (1 - p_\text{idle}) \cdot p_d + p_\text{idle} \cdot p_{f} \right]^K  \\ \nonumber
				&= \left[ p_d + (p_{f} - p_d) \cdot p_\text{idle}  \right]^K,
\end{align}
and where (a) becomes a lower bound when $M=1$ and $p_d = 1$ and $p_f =0$~\cite{Christensen:2010:NAF:1850837.1850860}.
From \eqref{eq:PhantomSignature}, the required signature frame size $\hat{L}$ to meet the target $\hat{p}_\text{fa}$ is
\begin{equation}\label{eq:LNonOptimal}
	\hat{L} = \frac{K}{M} \left[ 1 - \left(\frac{\hat{p}_\text{fa}^{1/K}-p_d}{p_f - p_d}\right)^{1/N} \right]^{-1}
\end{equation}
\begin{algorithm}[t]\label{alg:signatureDetection}
	\textbf{Input}: {$\mathbf{S}$, $\mathbf{y}$, $L$, $M$, $K$}; \\
	\textbf{Initialize}: $\mathbf{V} = \mathbf{S}$, $\mathbf{D} = \emptyset$\;
	\For{$ i : 1 \cdots L \, M$}{
		\For{$ \mathbf{s^{(h)}} \in \mathbf{V} \setminus \mathbf{D}$}{
			\If{$\mathbf{s^{(h)}}(1:i) \neq \mathbf{s^{(h)}}(1:i) \bigotimes \mathbf{y}(1:i)$}{
				$\mathbf{V} = \mathbf{V} \setminus \{\mathbf{s^{(h)}}\}$\;
			}
			\If{$ \left( \mathbf{V} \setminus \mathbf{s^{(h)}}(1:i ) \right) \bigotimes \mathbf{y}(1:i) \neq \mathbf{y}(1:i) $ }
			{
			$\mathbf{D} = \mathbf{D} \cup \{\mathbf{s^{(h)}}\}$\;
				Report to $\mathbf{u^{(h)}}$ that $\mathbf{s^{(h)}}$ is decoded\;}
			
		}
		
	}
	\For{$ \mathbf{s^{(h)}} \in \mathbf{V} \setminus \mathbf{D}$}{
		$\mathbf{D} = \mathbf{D} \cup \{\mathbf{s^{(h)}}\}$;
		Report to $\mathbf{u^{(h)}}$ that $\mathbf{s^{(h)}}$ is decoded\;
	}
	\caption{Iterative signature decoding where $\mathbf{S}$ is the set of signatures and $\mathbf{D}$ is the set of decoded signatures.}
\end{algorithm}
To compute the $K$ that minimizes $\hat{L}$ in \eqref{eq:LNonOptimal}, we assume $p_d = 1$ and $p_f =0$.
Then, for a given $N$ and $L$, the value of $K$ that minimizes $p_\text{fa}$ is given by~\cite{Mitzenmacher2001}
\begin{equation}\label{eq:OptimalK}
	K_{\min} = \frac{L \cdot M}{N}  \ln 2
\end{equation}
We use \eqref{eq:OptimalK} to find the minimal required $\hat{L}$ via \eqref{eq:LNonOptimal}.
Furthermore, recall that each device can only activate up to a single preamble per RAO, resulting in the constraint
 \begin{align}
 K_{\min} = L \, \min \left(1,\frac{M}{N} \ln 2\right),
 \end{align} 
where we assume to work in the regime in which $\frac{M}{N} \ln 2 < 1$, i.e., where $N > M \ln 2$.
Now, the minimum $\hat{L}$ can be obtained by solving iteratively the following fixed-point equation obtained from combining \eqref{eq:LNonOptimal} and \eqref{eq:OptimalK}
\begin{equation}\label{eq:IterativeL}
	\hat{L} = \ceil[\Bigg]{ \frac{\ceil{K_{\min}}}{M} \left[ 1 - \left( \frac{p_{fa}^{1 / \ceil{K_{\min}}} - p_d}{p_f - p_d} \right)^{1/N}\right]^{-1}},
\end{equation}
which converges for $p_d \geq 0.99$ and $p_f \leq 10^{-3}$, i.e., the prescribed preamble detection performance~\cite{3GPPTS36.141}.


\subsection{Signature Decoding} 
\label{sub:receiver_performance}

A straightforward approach for signature decoding is to perform it after all RAOs of the signature frame have been received, i.e., after the BS has observed the whole signature frame.
An alternative is to perform the decoding iteratively after every received signature RAO, i.e., the BS attempts to decode a signature while only having access to a partial observation of the signature frame.
The latter strategy is inspired with the fact that $K$ active RAOs constituting a signature are randomly spread over the signature frame and, in principle, the BS does not have to wait until the end of the frame to detect a signature.
The decoding performance is the same for both strategies when all $L$ RAOs in the signature frame have been received, but the average latency in the latter approach is lower.
We provide in Alg.~\ref{alg:signatureDetection} an algorithmic description of the iterative signature decoding, where the notation $\mathbf{z}(1:i)$ corresponds to the first $i$ entries of vector $\mathbf{z}$.
The key steps of the Alg.~\ref{alg:signatureDetection} are steps 5 and 7.
In particular, in step 5 the BS discards the signatures that could not have generated the partial observation $\mathbf{y}(1:i)$ from the set of potentially active signatures $V$.
Obviously, it is expected that $V$ will decrease with the additional received RAOs.
In step 7, the BS detects the signatures whose combinations of active RAOs and preambles are uniquely contributing to the partial observation $\mathbf{y}(1:i)$.
Then the BS reports to the respective device that its signature has been decoded, which in the LTE-A protocol realization would correspond to the RRC Connection Setup message, as shown in Fig.~\ref{fig:ARPComparison}(b).
Finally, in steps 10--12, when all RAOs have been received, the BS reports all the signatures within the set $\mathbf{V} \setminus \mathbf{D}$ as decoded.
\begin{figure}[t]
	\centering
	\includegraphics[width=0.85\linewidth]{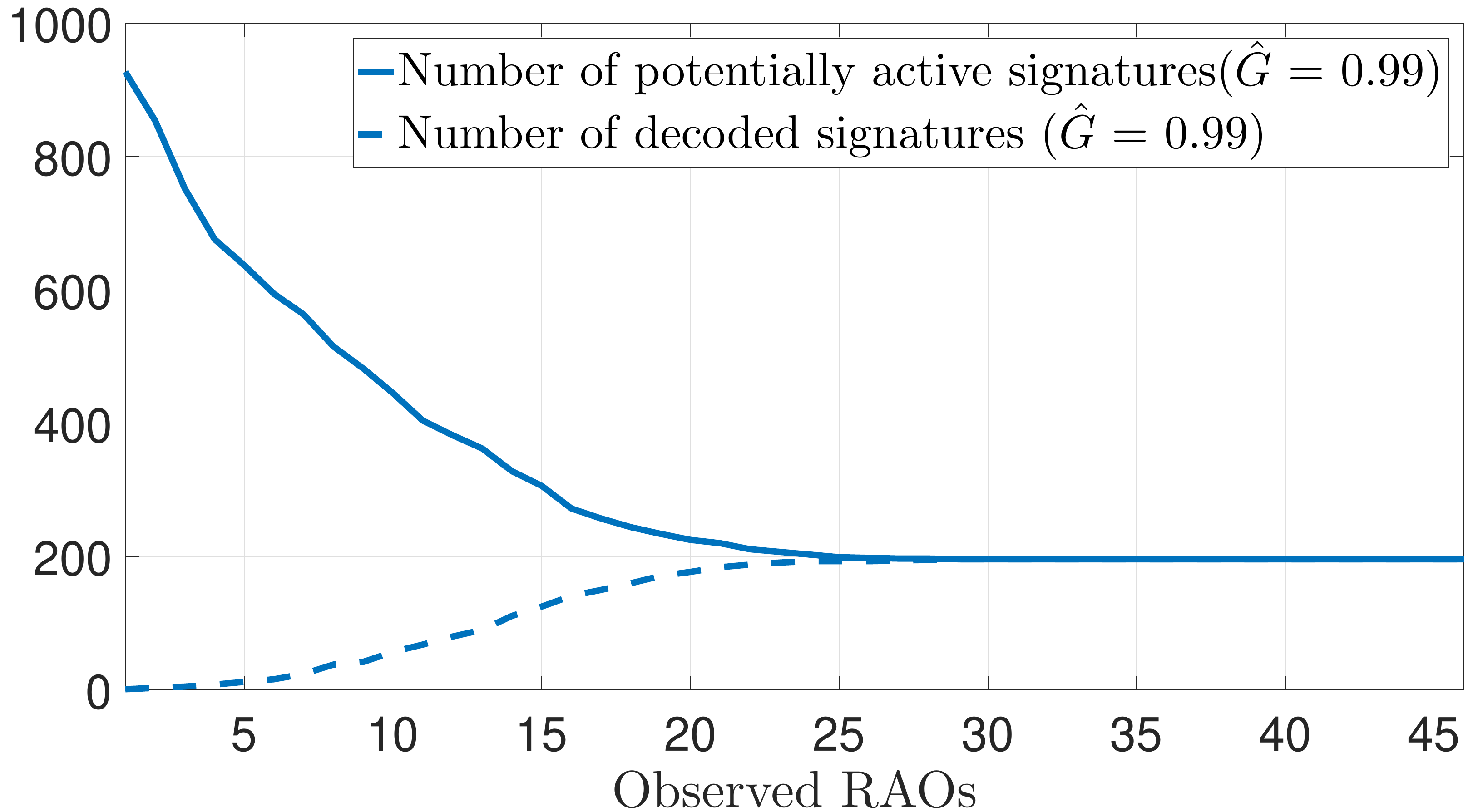}
	\caption{Evolution of the number of potentially active and already decoded signatures by the BS as the RAOs of the signature frame elapse, for $T=1000$, $N = 200$, $\hat{G}=0.99$, $p_d = 0.99$, $p_f = 10^{-3}$, and $\hat{L}=47$ from~\eqref{eq:IterativeL}.}
	\label{fig:advancedReceiverTrace}
	\vspace{-0.4cm}
\end{figure}

In Fig.~\ref{fig:advancedReceiverTrace}, we provide a simulation snapshot showing how many signatures are considered potentially active and how many have actually been decoded as the RAOs of the signature frame elapse.
Obviously, the iterative signature decoding occurs in a spread manner, which leads to the spreading of the feedback messages acknowledging the decoding of each signature, i.e., the RRC Connection Setup message in Fig.~\ref{fig:ARPComparison}(b).
In this way, the scenario in which a high number of devices attempt to complete the access reservation protocol simultaneously is avoided, i.e., the occurrence of congestion at the later stages of the ARP is reduced.
Another important observation is that most of the signatures become decoded well before the end of the signature frame.



\section{Performance Evaluation} 
\label{sec:system_performance_evaluation}

\subsection{Scenario description} 
\label{sub:scenario_description}

In order to evaluate the performance of the proposed signature based access and compare it with the proposed 3GPP LTE-A solution for MTC traffic~\cite{3GPPTR37.869}, we have implemented an event driven simulator where the main downlink and uplink LTE channels are modeled.
Specifically, the simulator implements the both procedures depicted in Fig.~\ref{fig:ARPComparison}(a) and Fig.~\ref{fig:ARPComparison}(b), while the downlink control and data channels (PDCCH and PDSCH respectively) and the uplink data and random access channels (PUSCH and PRACH) are modeled as in \cite{3GPPTR37.869}.

We consider a typical cell, configured with one RAO every 1~ms, $M=54$ available preambles for contention~\cite{3GPPTR37.869}.
We assume a total population of size $T = 1000$, and a batch arrival of $N_a$ devices with a payload of $100$ bytes to transmit,
The arrival probability of an individual device is given by $p_a = N/T$, i.e., $N_a$ is a binomially distributed random variable with mean $\text{E} [ N_a ] = N$.
The mean number of arrivals $N$ is assumed to be known, and the signature based scheme is dimensioned for it.\footnote{$N$ can be estimated, e.g., using techniques that take advantage of the LTE-A ARP, such as the one proposed in~\cite{MassiveM2MAccessWithReliabilityGuaranteesInLTESystems}.}
The probability of preamble detection by the BS is set to $p_d = 0.99$ and the probability of false detection of a preamble is set to $p_f = 10^{-3}$ \cite{3GPPTS36.141}.

In the baseline, i.e., 3GPP scheme, we assume the typical values for the backoff window of 20~ms and the maximum number of $10$ connection attempts~\cite{3GPPTR37.869}.
The devices upon becoming active contend for access by activating randomly one preamble in one of the available RAOs within the backoff interval, i.e., the batch arrival is spread with the backoff interval.\footnote{Note that this initial backoff is a modification of the original LTE-A access procedure, in which the devices contend by activating a preamble in the nearest RAO~\cite{3GPPTR37.868}. The purpose of this modification is to force a spread in the batch arrival and prevent the consequent imminent collision; the resulting performance of the baseline scheme is actually better than it could be expected.}
In case that a device is the only one that selected a given preamble in a given RAO and that this preamble has been detected, then the access procedure, as depicted in Fig~\ref{fig:ARPComparison}(a), proceeds until completion. Otherwise, the device will reattempt the access within the back-off window after the timer to receive the RAR as elapsed.
When multiple devices select the same preamble within a RAO, the resources assigned by the BS corresponding to the step 3 in the protocol are wasted due to the collided devices; and the collided devices re-attempt access later by selecting a random RAO within the backoff interval.
The devices re-attempt access until either successful or until exceeding the allowed number of retransmissions.

In the proposed method, the devices contend by transmitting their signatures, where the signature frame length $L$ is obtained from~\eqref{eq:IterativeL}.
For the sake of comparison, we also evaluate the performance of the random signature construction~\cite{ETT:ETT2656}, where $K = L$.
Each device upon its signature being decoded, even in the case of false positive, receives the feedback RRC connection setup message and is assigned uplink data resources for the transmission of the third and final message, see Fig~\ref{fig:ARPComparison}(b).

The performance is evaluated in terms of: (i) the average goodput $E[G]$; (ii) the average latency until the first step in both access schemes is successful,  corresponding to a singleton preamble in the baseline and a successfully decoded signature in the proposed scheme; (iii) the average latency until the small data transmission takes place, corresponding to step 5 in the baseline and to step 3 in the proposed scheme, see Fig~\ref{fig:ARPComparison}; and (iv) probability of device being successfully detected upon the completion of the access protocol.

The average goodput $E[G]$ is evaluated as the ratio between the successfully used resources and the total resources spent in the third step of both access protocols.
It directly relates to the efficient use of resources, since the BS is only able to discern if there is a correctly detected device upon successful completion of the third step.
In the baseline scheme, the system resources are wasted whenever two or more devices select the same preamble within a RAO; the goodput in this case is given as the ratio between the total number of messages that are exchanged successfully and the total number of exchanged messages at the third step, including the failed ones due to collisions. 
In the case of the signature based access, the wasted resources in the third step occur whenever a false positive signature occurs, and the goodput is given by~\eqref{eq:G_def}.

\begin{figure}[tb]
	\centering
		\includegraphics[width=0.95\linewidth]{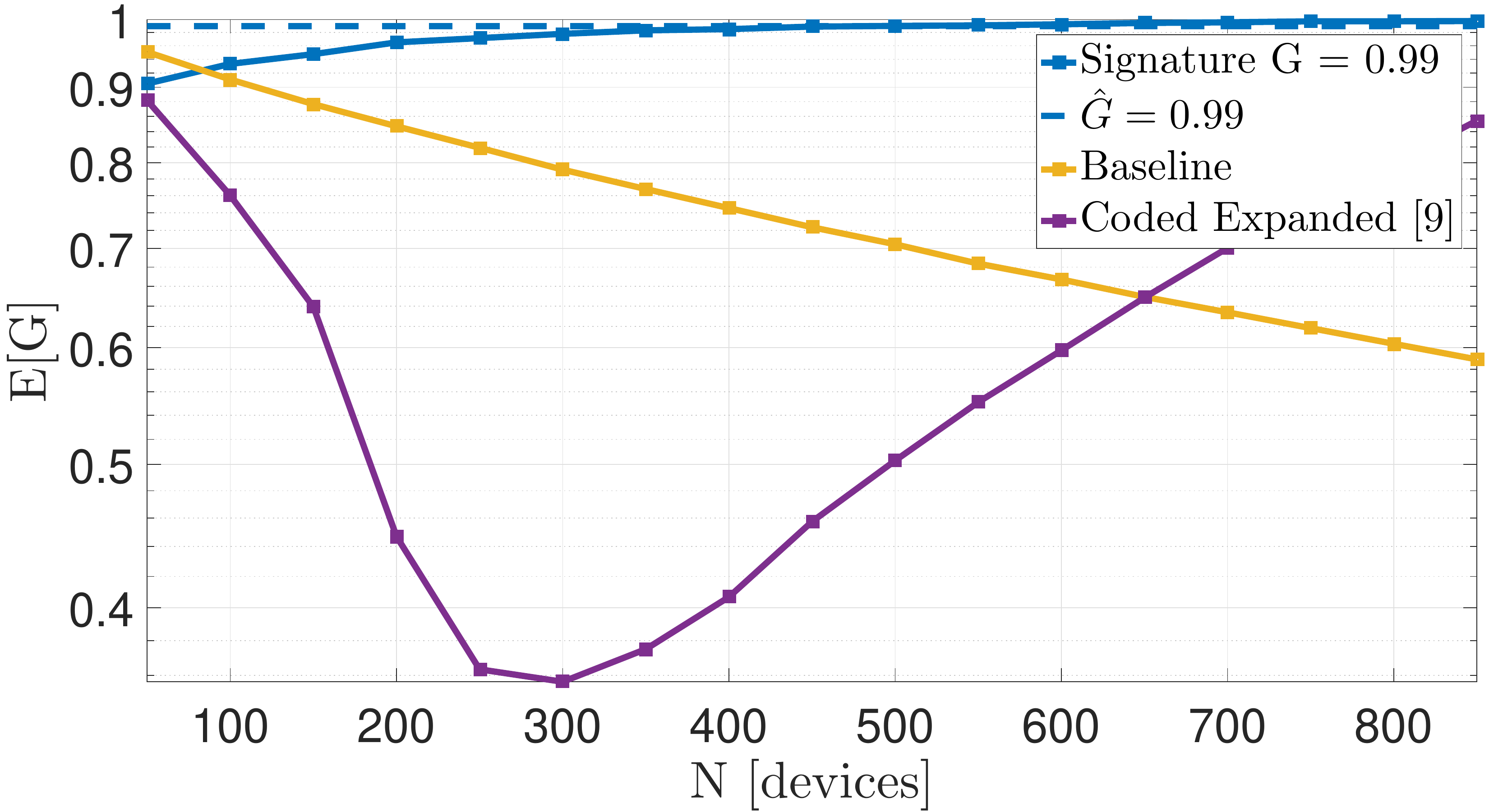}
	\vspace{-0.1cm}
	\caption{$E[G]$ observed with increasing $N$, for the 3GPP scheme, random signature construction~\cite{ETT:ETT2656} and the proposed signature construction. ($T=1000$)}
	\vspace{-0.5cm}
	\label{fig:Goodput}
\end{figure}

\subsection{Results} 
\label{sub:numerical_results_and_discussion}

The expected goodput is depicted in Fig.~\ref{fig:Goodput}, where for the goodput target for the proposed method~\eqref{eq:phantomTarget} is set to $\hat{G} = 0.99$.
We observe that the proposed method meets the actual goodput meets the design target at higher access loads. On the other hand, at lower $N$, the performance deviates from the target value $\hat{G} = 0.99$. This is due to the assumption that the false positive signatures are independently and uniformly generated from the idle signatures, which is the basis of the approximation in \eqref{eq:PhantomSignature}.
We can also observe that the goodput performance of the proposed method is always superior to the 3GPP scheme.
Specifically, In the 3GPP scheme the devices re-attempt retransmission upon colliding and until they are either successful or the number of retransmissions is exceeded.
Each subsequent failed retransmission results in additional wasted system resources, which results in the observed degradation of the baseline goodput with increasing number of active devices.
Finally, the goodput achieved with the random signature construction \cite{ETT:ETT2656} is quite low, due to the high number of false positives. 
\begin{figure}[tb]
	\centering
		\includegraphics[width=0.95\linewidth]{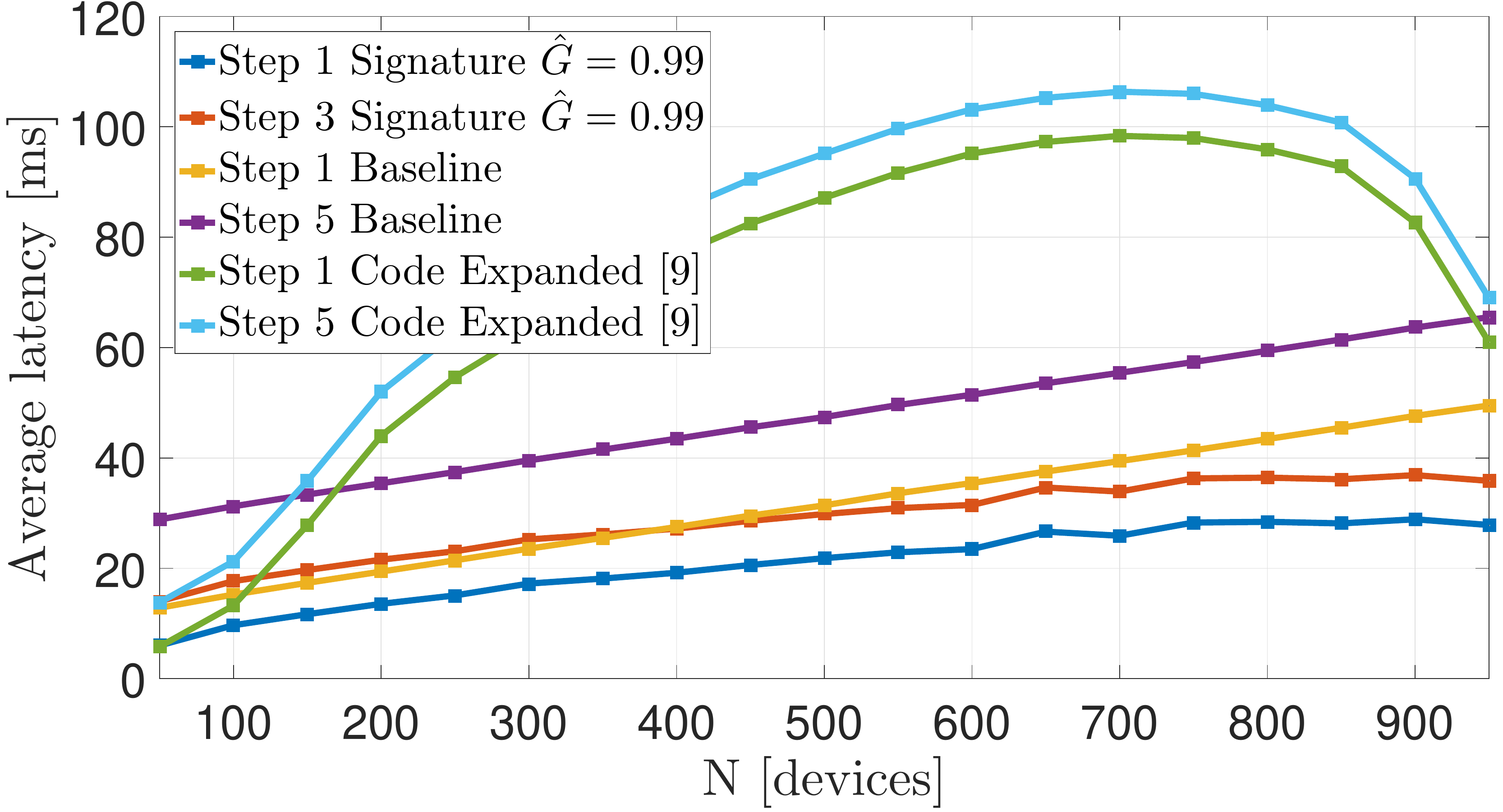}
	\vspace{-0.1cm}
	\caption{Mean latency of the 3GPP scheme, random signature construction and the proposed signature construction with optimal $K$ and minimum $\hat{L}$ computed from~\eqref{eq:IterativeL}, at different stages of the access procedures. ($T=1000$)}
	\vspace{-0.5cm}
	\label{fig:Latency}
\end{figure}

In Fig.~\ref{fig:Latency} we depict the mean latency at step 1 in all schemes, as well as in steps 3 and 5 in the signature and 3GPP schemes, respectively.
An important observation is that the latency of the proposed method is always lower than the 3GPP scheme; and the gap between these two schemes increases for higher $N$.
This is a consequence of the more efficient detection of active users, as can be seen when comparing the latency of these two schemes at step 1.
Furthermore, the random signature construction has the worst performance, the reason being that a signature cannot be decoded before all $L$ RAOs of the signature frame have been received~\cite{ETT:ETT2656}.

Finally, in Tab.~\ref{table:ProbDetectionTable} we show the probability of a device being successfully detected at end of the access protocol.
Here the proposed method has a slight performance degradation compared to the 3GPP scheme, but this degradation diminishes higher access loads.
The 3GPP scheme achieves higher detection performance due to only requiring one transmission out of all preamble retransmissions to be successful, making it more robust but at the cost of lower goodput and higher latency. 
On the other hand, the random signature construction leads to a very low detection performance, as it requires the successful detection of all the active preambles~\cite{ETT:ETT2656}.
%


\section{Discussion and Conclusions} 
\label{sec:conclusions}

Following the insights provided by Bloom filters, we have introduced the concept of signatures with probabilistic guarantees and applied it to a system model derived from the LTE-A access reservation protocol.
The most important feature of the proposed method is in allowing the device to be identified already at the access stage.
Moreover, the method is very efficient in terms of use of the system resources and has a favorable performance in terms of decoding latency.

In the paper we assumed that the base station serves the successfully connected devices without preferences.
Nevertheless, it is straightforward to modify the proposed solution to scenarios in which the BS serves devices based on the identifications inferred from the decoded signatures, i.e.,  IMSIs and/or connection establishment causes.
In such cases, the proposed access method enables differentiated treatment by the BS from the very beginning.

Finally, we note that in the paper we assessed a simplified scenario of a synchronous bath arrival in order to present the key concepts and the related analysis.
Tuning the proposed scheme for the other typical models, like the Beta arrival model for synchronous arrivals or the Poisson arrival model for asynchronous arrivals, is left for further work.


\begin{table}[t]                                                           
\centering                                                                  
\begin{tabular}{  c  c c c c c  }                                      
\hline                                                                      
\textbf{N} & 100 & 300 & 500 & 700 & 900 \\
\hline      
Proposed method & 96 & 98 & 98 & 98 & 98 \\                                                          
3GPP scheme & 100 & 100 & 100 & 100 & 100  \\        
Random construction \cite{ETT:ETT2656} & 86 & 53 & 42 & 37 & 44 \\       
\hline                                                                      
\end{tabular}
\vspace{0.1cm}                                                               
\caption{Probability of successfully detecting a device [\%]. (T = 1000)}
\vspace{-0.9cm}
\label{table:ProbDetectionTable}                                                
\end{table}

\section*{Acknowledgment}
This work was performed partly in the framework of H2020 project FANTASTIC-5G (ICT-671660), partly supported by the Danish Council for Independent Research grant no. DFF-4005-00281 ``Evolving wireless cellular systems for smart grid communications'' and by the European Research Council (ERC Consolidator Grant Nr. 648382 WILLOW) within the Horizon 2020 Program.
The authors acknowledge the contributions of the colleagues in FANTASTIC-5G.

\end{document}